\documentclass[%
 reprint,
superscriptaddress,
 amsmath,amssymb,
 aps,
]{revtex4-2}

\usepackage{graphicx}
\usepackage{dcolumn}
\usepackage{bm}
\usepackage{xspace}
\usepackage{xcolor}
\usepackage{subcaption}
\usepackage{ragged2e}
\usepackage{siunitx}

\newcommand{\au}{\,\mathrm{a.u.}}
\newcommand{\eV}{\,\mathrm{eV}}

\newcommand{\Wcm}{\,\mathrm{W/cm^2}}
\newcommand{\HHO}{H$_2$O\xspace}
\newcommand{\Efield}{\mathcal{E}}

\begin{document}

\preprint{APS/123-QED}

\title{Intensity-resolved measurement of above-threshold ionization of Ar-H$_2$O}

\author{Adrian Platz}%
\affiliation{Institute for Optics and Quantum Electronics, Universit\"at Jena, D-07743 Jena, Germany}
\author{Sebastian Hell}
\affiliation{Institute for Optics and Quantum Electronics, Universit\"at Jena, D-07743 Jena, Germany}
\author{Yinyu Zhang}
\affiliation{Institute for Optics and Quantum Electronics, Universit\"at Jena, D-07743 Jena, Germany}
\author{Bo Ying}
\affiliation{Institute for Optics and Quantum Electronics, Universit\"at Jena, D-07743 Jena, Germany}
\affiliation{Helmholtz Institute Jena, D-07743 Jena, Germany}

\author{Gerhard G. Paulus}
\affiliation{Institute for Optics and Quantum Electronics, Universit\"at Jena, D-07743 Jena, Germany}
\affiliation{Helmholtz Institute Jena, D-07743 Jena, Germany}

\author{Matthias K\"ubel}
\affiliation{Institute for Optics and Quantum Electronics, Universit\"at Jena, D-07743 Jena, Germany}
\affiliation{Helmholtz Institute Jena, D-07743 Jena, Germany}
\email{matthias.kuebel@uni-jena.de}
%


\date{\today}

\begin{abstract}
Above-treshold ionization (ATI) by femtosecond laser pulses centered at 515\,nm is studied for a gas mixture containing the Van-der-Waals complex Ar-H$_2$O. By detecting photoions and -electrons in coincidence, the ATI spectra for Ar, Ar$_2$, \HHO, and Ar-\HHO are discerned and measured simultaneously. Using an intensity-scanning technique, we observe the red-shift of the ATI spectra as a function of the laser intensity. The intensity-dependent shift of the ATI peak positions observed for Ar-H$_2$O and H$_2$O match but significantly differ from the ones measured for Ar and Ar$_2$. This indicates that the photoelectron is emitted from the \HHO site of the complex and the vertical ionization potential of Ar-H$_2$O is determined as $(12.4 \pm 0.1)$\,eV. For resacttered electrons, however, an enhancement of high-order ATI is observed for Ar-H$_2$O, as compared to H$_2$O, suggesting that the relatively large Ar atom acts as a scattering center, which influences the ionization dynamics. 
\end{abstract}

\maketitle


\section{Introduction}
Rare-gas compounds exhibit rich light-matter interactions, and serve as model systems for electron dynamics in weakly bound molecules. A prominent example is interatomic Coulombic decay, which has been first observed in Ne$_2$ and studied using extreme ultraviolet radiation \cite{Jahnke2020,Trinter2022}. Also in intense optical fields, rare gas dimers have attracted significant interest \cite{Wunderlich1997, Kunitski2019, Tong2022} but only few studies have explored light-matter interaction in more complex rare-gas compounds \cite{Wu2012}. A notable example for such complex target is Ar-\HHO, which has been extensively studied in the past using infrared vibrational spectroscopy \cite{Cohen1993,Germann1993,Tao1994,Weida1997,Liu2014}. One focus of attention has been the location of the energy minimum in the Ar-\HHO intermolecular potential energy surface. After initially contradicting results, studies have converged to the finding that the argon atom resides in the plane of the water molecule, at an angle close to one of the hydrogen atoms but at roughly four times the distance from the oxygen atom \cite{Cohen1993,Tao1994,Makarewicz2008,Hou2016}. Despite this progress, ionization experiments of Ar-\HHO have been scarce. Conseqeuently, little is known about the structure of the Ar-\HHO cation. 

Structural and dynamical information from atoms and molecules has been successfully infered from above-threshold ionization (ATI) spectra, for example using laser-induced electron diffraction \cite{Meckel2008,Blaga2012, Wolter2016}, or photoelectron momentum imaging \cite{Eckart2018, Kunitski2019, Kuebel2019}. For recent reviews see \cite{DeGiovannini2023,Amini2021}. Moreover, spectroscopy of electronic states in polyatomic molecules has been demonstrated in channel-resolved ATI experiments \cite{Boguslavskiy2012,Mikosch2013}. However, accurately measuring the ionization potential using ATI is challenging since the ATI peak positions sensitively depend on the laser intensity. The position of the S-th ATI peak can be described by (atomic units are used if not otherwise stated)
\begin{equation}
    E_S = (N+S) \omega - E_\mathrm{IP} - \Delta E(I), \label{eq:ATI_peak}
\end{equation} 
where $\omega$ is the photon energy, $E_\mathrm{IP}$ is the ionization potential of the target atom or molecule, $N = \lceil E_\mathrm{IP} / \omega \rceil$ is the minimum number of photons required to overcome $E_\mathrm{IP}$, and $\Delta E(I)$ is an intensity-dependent peak shift. The latter is often approximated by the ponderomotive potential $U_P = \frac{I}{4\omega^{2}}$, representing the AC Stark shift of the ionization continuum. In addition, the ground state exhibits an intensity-dependent level shift, as well. In the absence of resonances and for not too intense fields, the AC Stark shift can be reasonably well approximated by the DC Stark shift, calculated by second-order perturbation theory \cite{Delone1999}. It is given in terms of the peak intensity $I = \Efield_0^2$, where $\Efield_0$ is the peak value of the electric field,  and the ground state polarizability $\alpha$ of the respective atom or molecule, by
\begin{equation}
\delta E = -\alpha  I/4, \label{eq:stark_ground_state}
\end{equation} 
and is usually negative.
Hence, the ground state stark shift usually increases the peak shift $\Delta E(I)$. Note that this does not hold in the presence of resonances. For example, Nicklich, \emph{et al.} observed an intensity-dependent blue-shift in the multiphoton ionization of Cs by visible wavelengths \cite{Nicklich1992}.
The intensity-dependent peak shift of ATI has been observed in various experiments, which aimed at reducing the focal volume averaging effect \cite{Walker1998,Hart2014,Wiese2019}, or identifying the role of resonances \cite{Wiese2019,Wang2014}. In many of these experiments, the Stark shift of the ground state has been neglected. 

Here, we present intensity-resolved measurements of ATI of Ar, Ar$_2$, \HHO, and Ar-\HHO by 515\,nm femtosecond laser pulses. The data allows us to extract the unkown ionization potential of Ar-\HHO and infer that the electron is emitted from the \HHO site of the compound. In contrast to this, we show that elastic electron rescattering also occurs on the argon atom, whose presence significantly enhances the elastic scattering cross-section of Ar-\HHO with respect to \HHO. 

\section{Experimental approach}
In our experiment, the gas sample containing Ar-\HHO is obtained by bubbling argon at a pressure of 3 bar through water and co-expanding the mixture through a 30-$\mu$m-nozzle into a high-vacuum vessel. The nozzle was heated to 50°C to avoid clogging by condensed water. The gas jet subsequently passes a 150-$\mu$m-skimmer and two 1-mm apertures for differential pumping before it enters a Cold Target Recoil Ion Momentum Specrometer (COLTRIMS), where it intersects the laser focus at a total distance of $\approx \SI{1.5}{m}$ behind the nozzle. 

\begin{figure}
  	 	\includegraphics[trim={5cm 0cm 5.5cm 1cm},clip,width = 0.38\textwidth]{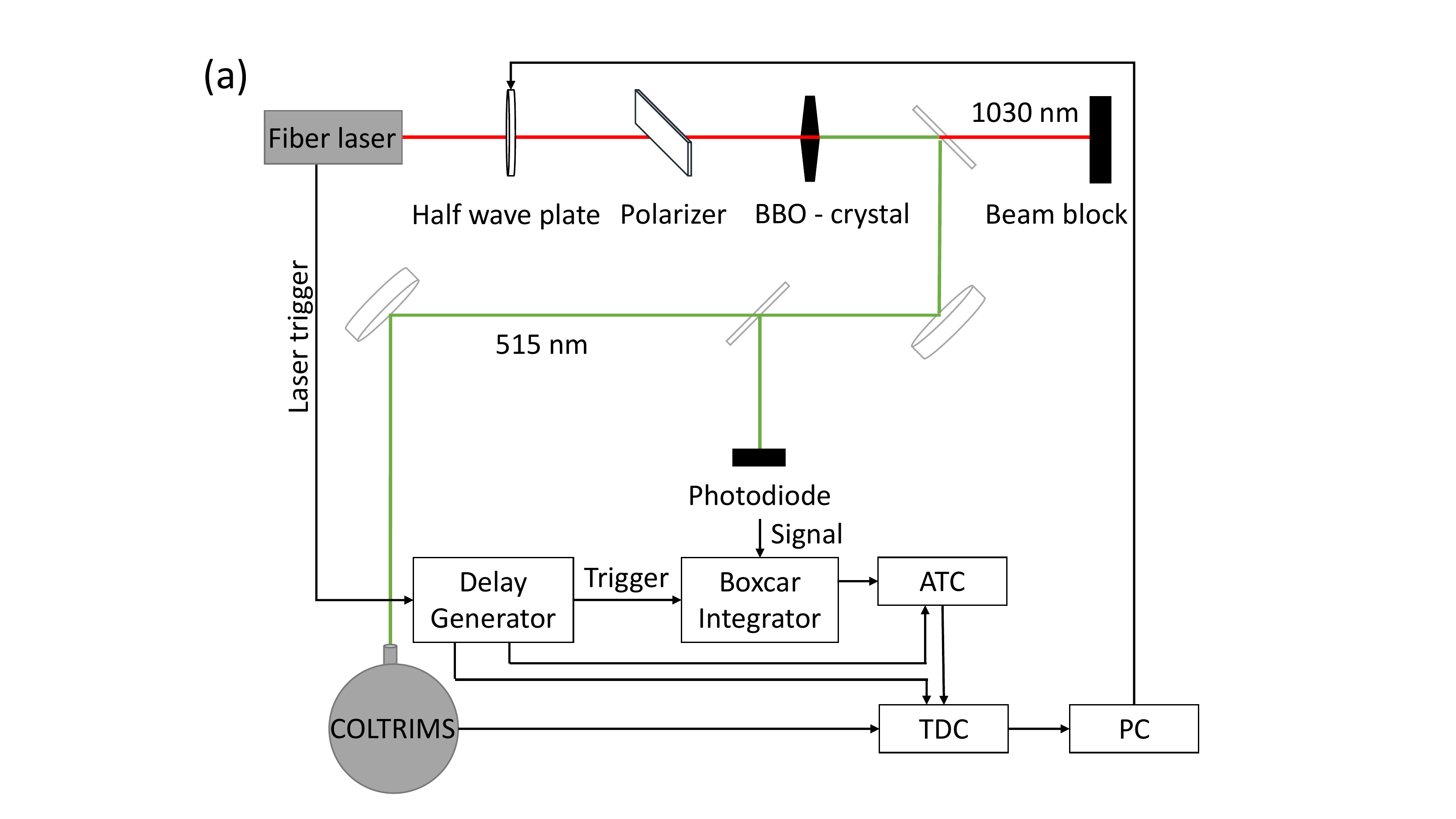}
    \hfill
    	\includegraphics[width = 0.45\textwidth]{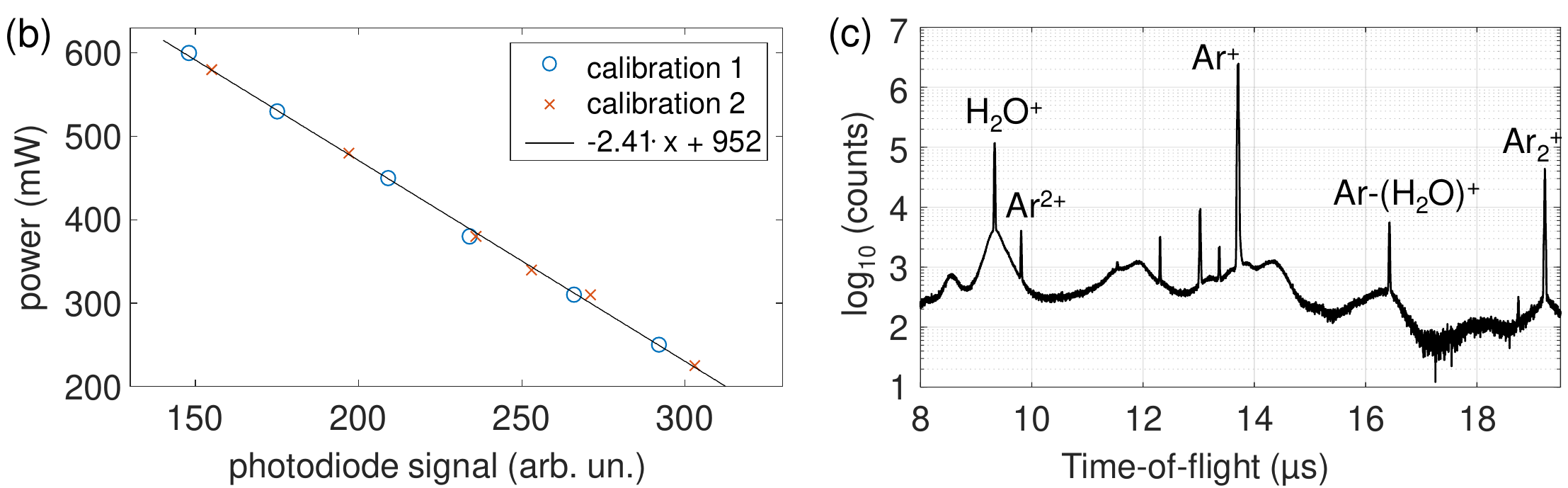}
    \caption{\small \justifying
        (a) Schematic of the experiment for intensity-resolved coincidence measurements of above-threshold ionization. The power of the femtosecond laser pulses at 1030\,nm with are controlled using a motorized half-wave plate followed by a polarizer. The laser frequency is doubled in a BBO, and the two colours are separated. The laser power at 515\,nm is measured by a photodiode and a boxcar integrator. In the data acquisition computer, the measured laser power is correlated with the ion and electron momenta detected in the COLTRIMS. (b) Recorded photodiode signal as a function of the laser power, as measured with a power meter before (calibration 1) and after (calibration 2) conducting the main experiment. (c) Ion time-of-flight spectra recorded in the COLTRIMS, with various ionic species marked. }
    \label{fig1:schematic}
\end{figure}

Figure \ref{fig1:schematic}(a) shows a schematic of the experiment. Femtosecond laser pulses centered at 1030$\,$nm 
are obtained from a commercial Yb-doped fiber laser (Active Fiber Systems) operated at 50$\,$kHz repetition rate. The pulses are compressed to 36$\,$fs using an argon-filled hollow-core fiber and suitable chirped mirrors. The laser power is controlled using a motorized half-wave plate and a broadband thin-film polarizer. The transmitted light is frequency-doubled using a 300-$\mu$m thick Beta Barium Borate (BBO) crystal and the fundamental infrared beam is subsequently discarded. A small fraction of the visible beam is reflected off a wedge and impinges on a photodiode, operated well below saturation. The photodiode signal is further processed using a boxcar integrator to yield a measurement of the pulse energy. Importantly, the photodiode signal is proportional to the laser power recorded with a power meter, as specifically shown for the range of 200 mW to 600 mW in Fig.~\ref{fig1:schematic}(b). Since the second harmonic generation was operated well below saturation, we assume that the pulse duration and beam profiles remains unchanged when the laser power is varied using the motorized half-wave plate. Thus, the photodiode signal is proportional to the intensity in the laser focus. The main beam is focused (7.5$\,$cm focal length) into the center of a Cold Target Recoil Ion Momentum Spectrometer (COLTRIMS) \cite{Ullrich2003} where it intersects a cold gas jet containing the argon-water gas mixture. The momenta of ions and electrons produced in the laser focus are measured in coincidence using time and position sensitive detectors on either side of the COLTRIMS. Figure \ref{fig1:schematic}(c) shows the ion time-of-flight spectrum, which permits the distinction of various ionic species by their time-of-flight, and the accurate measurement of the recoil momentum along the laser polarization. Momentum conservation between ion and photoelectron allows us reliably assign photoelectrons to the ion of the species from which they originate. A total of $6.4 \times 10^4$ counts are detected for the Ar-H$_2$O$^+$ + e$^-$ and $3.2\times 10^7$ counts for the Ar$^+$ + e$^-$ coincidence channels. 

Besides ion-electron coincidences, coincident events of two ionic species, notably, Ar$^+$ + H$_2$O$^+$ Coulomb explosion events are observed. The kinetic energy release of this break-up channel has a mean value of 3.820\,eV and a width (standard deviation) of 0.23\,eV.
Assuming instantaneous double ionization and subsequent propagation of a $1/r$ repulsive potential, this corresponds to a bond distance between Ar and \HHO of $(3.77 \pm 0.23)\,\AA$, in reasonable agreement with but slightly longer than the predicted values of $3.636\,\AA$ \cite{Cohen1993} or $3.603\,\AA$ \cite{Tao1994}. The experimental result might be systematically shifted towards larger values due to the nonlinear ionization rate, especially in the second ionization step, for which the ionization potential is expected to strongly decrease with increasing internuclear distance. Moreover, we note that the measured KER results from the effective distance of the two charges rather than the center-of-mass distance between Ar and \HHO.

\begin{figure}
    \centering
    \includegraphics[width = 0.5 \textwidth]{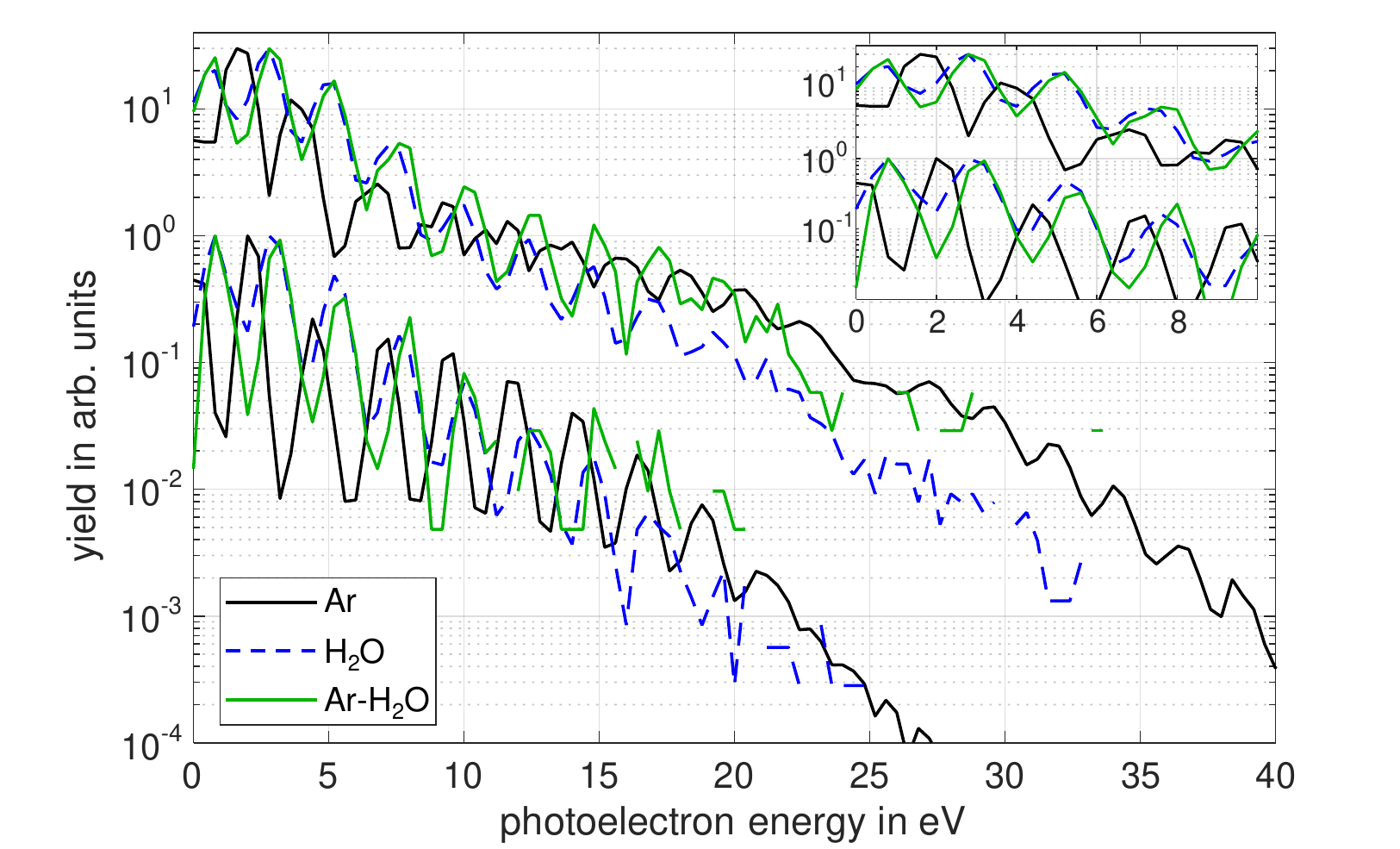}
    \caption{\small \justifying Measured Energy distributions of photoelectrons detected in coincidence with Ar$^+$ (black), H$_2$O$^+$s (dashed blue) and Ar-H$_2$O$^+$ (green). Each curve is normalized to its maximum (851059 counts for Ar, 22822 for \HHO and 1036 for Ar-\HHO). Missing data points are due to a lack of statistics. The upper curves represent all data acquired in the experiment with a peak intensity of $\approx 1.0\times 10^{14}\Wcm$, the lower ones were recorded at $ \approx 0.5\times 10^{14}\Wcm$. The data sets with different intensity values are vertically displaced for visual convenience. The inset shows the low-energy range of the ATI spectra.}
    \label{fig2:ATI}
\end{figure}

\section{Results and Discussion}
\subsection{Intensity-dependent ATI spectra}
Figure \ref{fig2:ATI} shows photoelectron energy spectra for ATI of Ar, \HHO, and Ar-\HHO. The spectra exhibit a typical shape with a rapid drop over the first few ATI orders, followed by the rescattering plateau \cite{Paulus1994} that cuts off around 30\,eV (20\,eV) for the upper (lower) curve. Remarkably, the signal in the plateau region is approximately two times higher for Ar and Ar-\HHO than for \HHO.  We will return to this observation in the context of the photoelectron momentum distributions presented in Fig.~\ref{fig6:momentum distributions} below. The photoelectron spectra are modulated by the ATI peaks, which are spaced by the photon energy of 2.4\,eV, as expected from equation \ref{eq:ATI_peak}. Furthermore, the ATI combs recorded for different targets are shifted with respect to each other, due to the difference in the ionization potentials. 
A double peak structure is observed for Ar at higher intensity, which reduces the contrast between ATI peaks and suggests the involvement of resonances. Comparing the ATI combs recorded at higher intensity (upper curves) to the ones recorded at lower intensity (lower curves) reveals the intensity-dependent peak shift. Importantly, a close examination of the peak positions (see inset of Fig.~\ref{fig2:ATI} shows that the offset between the peak positions observed for Ar and Ar-\HHO differ for the two intensity values. Hence, the ionization potential of Ar-\HHO cannot be retrieved simply by comparing the respective ATI peak positions to the ones measured for Ar. The ATI peaks measured for \HHO and Ar-\HHO, however, are close to each other for both intensity values. Whether the offset between the ATI peaks for the two targets allow us to extract the ionization potential of Ar-\HHO will be tested using an intensity scan.

\begin{figure}

\includegraphics[width = 0.5 \textwidth]{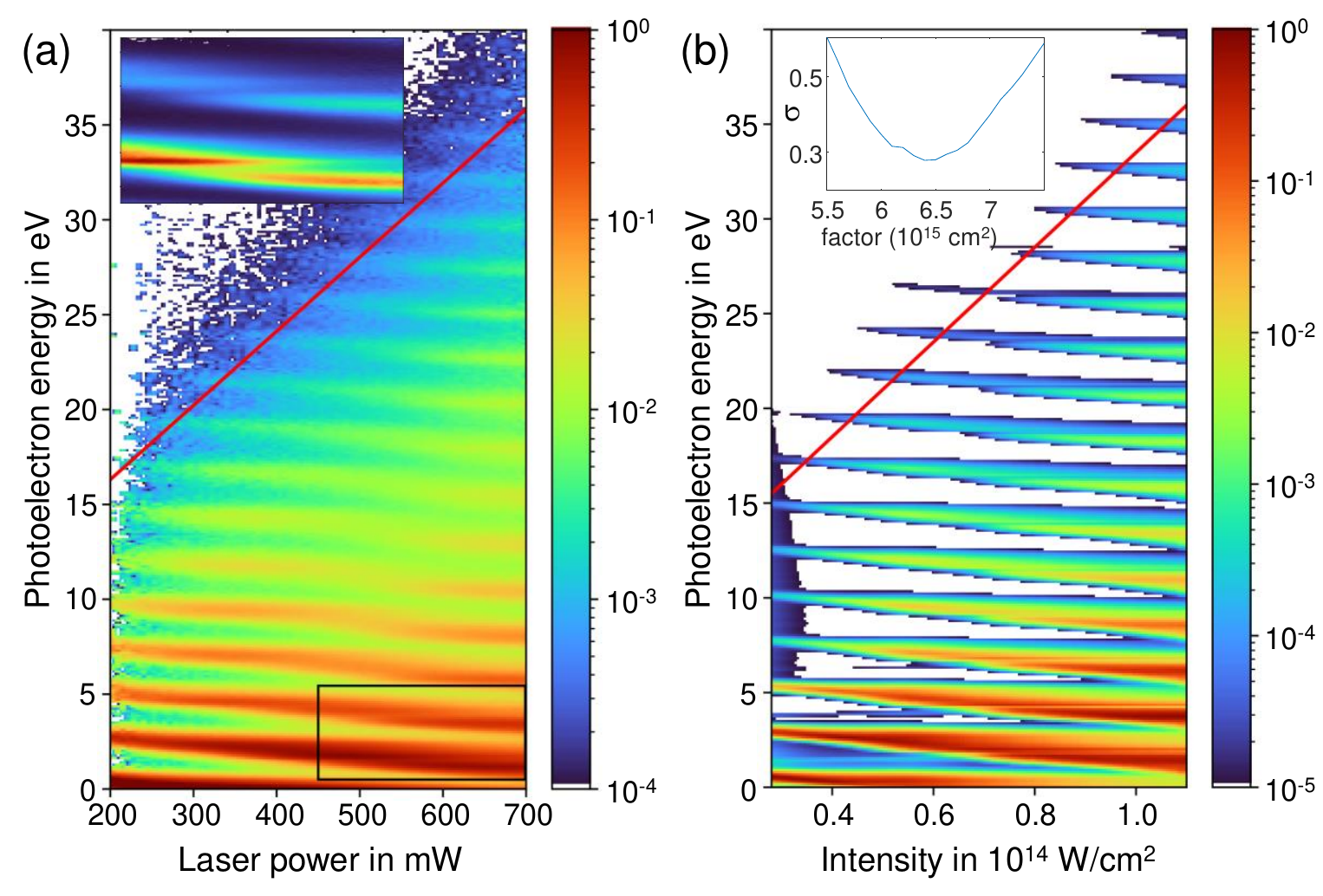}
  \caption{\small \justifying Intensity-dependent ATI spectra for ionization of Ar by 515\,nm light. (a) Experimental results obtained by integrating the measured photoelectron momentum distributions over the full solid angle. The red solid line indicates the cut-off energy of the photoelectron spectra and is described by the expression $0.0391 \eV / \mathrm{mW} + 8.5\eV$. (b) Numerical results obtained by solving the one dimensional TDSE. The red line is described by the expression $25.0 \eV / (10^{14}\Wcm) + 8.5\eV $ indicating the high-energy cut-off. The inset shows the  standard deviation $\sigma$ as a function of the proportionality constant between peak intensity and laser power. } 
  \label{fig3:ATI_intensity}
\end{figure}

The intensity scan is carried out by scanning the half-wave plate while the laser power transmitted trhough the polarizer is tracked by the photodiode, as described above. The measured power-dependent ATI spectra for Ar are presented in Fig.~\ref{fig3:ATI_intensity}(a). The ATI spectra exhibit a pronounced intensity dependence: while the cut-off energy increases proportionally to the laser power, the individual ATI peaks are red-shifted and also broadened. This behaviour is an impressive manifestation of the focal volume effect: with increasing peak intensity, more and more intensity values contribute to the ATI yield. Since different intensity values corresponds to different peak positions, the peaks in the observed ATI spectra are broadened and eventually washed out. In addition, Freeman (multi-photon) resonances \cite{Freeman1987,Cormier2003} lead to horizontal features in the intensity-dependent ATI spectrum (see inset of Fig.~\ref{fig3:ATI_intensity}), and are likely responsible for the double peak structure observed for Ar in Fig.~\ref{fig2:ATI} at higher intensity.

The intensity-resolved ATI spectra allow us to find the proportionality constant between measured laser power and peak intensity. This is achieved by comparison of the experimental results to numerical results displayed in Fig.~\ref{fig3:ATI_intensity}(b). These were obtained by solving the one-dimensional time-dependent Schr\"odinger equation \textbf{(TDSE)} using the split-operator method. The Hamiltonian is given by $H(x,t) = \frac{Z}{|x|+\alpha} + \Efield(t)\cdot x$. The softcore parameters $Z$ and $\alpha$ where tuned such that the energies of the ground state and first-excited state of the model atom match those of real argon. The laser field $\Efield(t) = \Efield_0(t) \cos{\omega t}$ has a Gaussian pulse envelope $\Efield_0(t)$ with a full-width at half-maximum of \SI{40}{fs} and a carrier frequency of $\omega = \SI{0.088}{a.u.}$
The calculated photoelectron spectra were integrated over a Gaussian focal volume \cite{Zhang2020} and agree well with the measured ones. The proportionality constant between peak intensity and laser power in the experimental data is determined by minimizing the standard deviation between the positions of ATI peaks of orders 1 to 10 in the measured and calculated ATI spectra. 
We find best agreement for $(6.40 \pm 0.30)\,\frac{\mathrm{mW}}{\mathrm{TW/cm^2}}$. The uncertainty indicates the range in which the standard deviation increases by no more than 10\%, see inset of Fig.~\ref{fig3:ATI_intensity}(b). Using this conversion, we confirm that the high-energy cut-off of the spectra can be adequately described by the extended cut-off law proposed in Ref.~\cite{Busuladzic2006}: $E_\mathrm{max} = 10\,U_P + 0.538 E_\mathrm{IP}$, as indicated by the red lines in Fig.~\ref{fig3:ATI_intensity}. 

\subsection{Peak shifts and ionization potential of Ar-\HHO}

\begin{figure}
    \centering
    \includegraphics[width = 0.5 \textwidth]{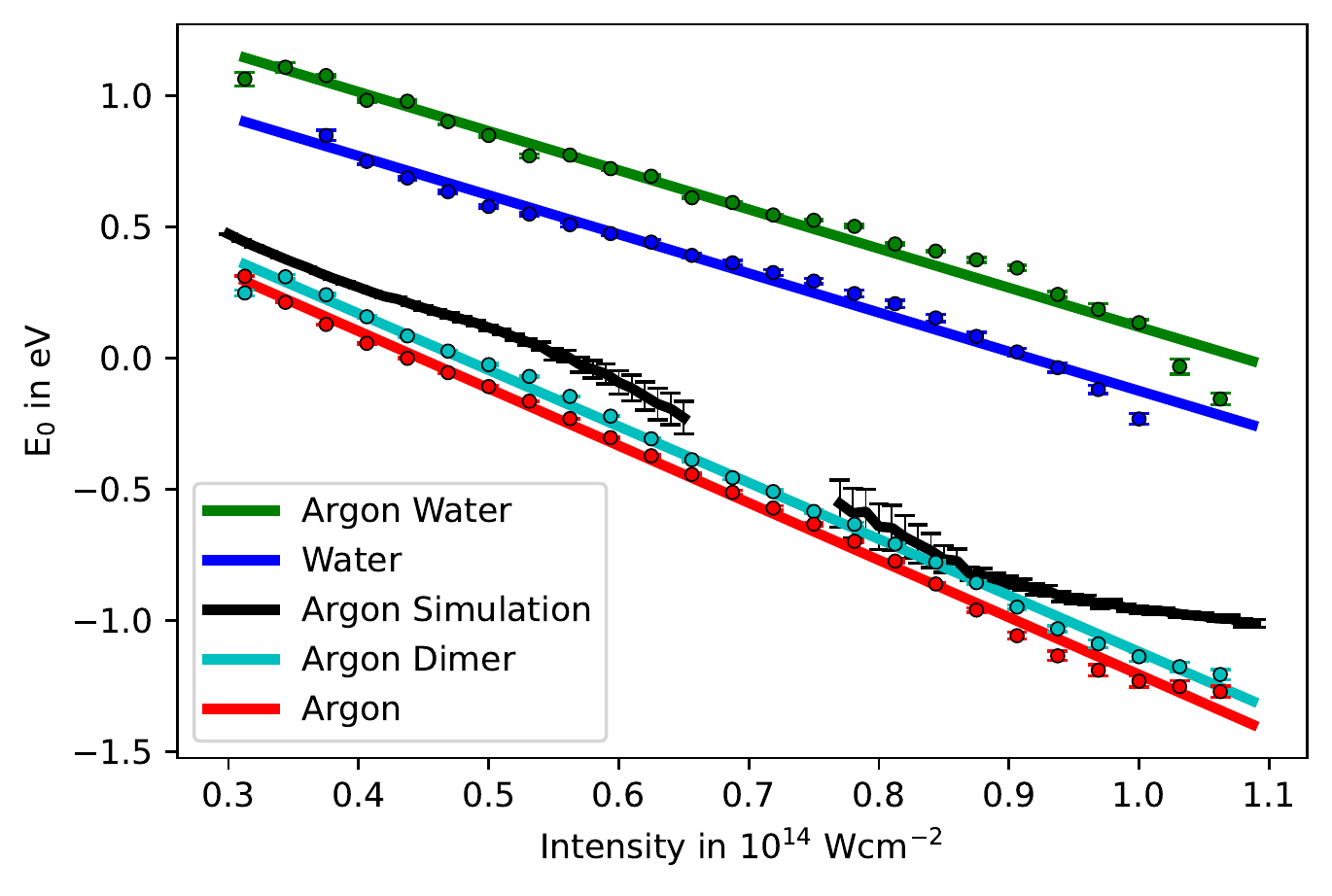}
    \caption{\small \justifying Intensity-dependent ATI peak positions for various targets. Shown is the measured offset energy of the ATI comb for Ar, Ar$_2$, \HHO and Ar-\HHO, as function of the peak intensity. The colored solid lines represent linear fit functions to the data. The peak positions extracted for the simulated data is also shown (black). Around $7 \cdot 10^{13} \Wcm$, no accurate fit could be obtained due to strong modulations in the calculated photoelectron spectra.}
    \label{fig4:peakshift}
\end{figure}

In order to quantitatively analyze the intensity-dependent ATI spectra for all targets, we extract the ATI peak positions from the measured photoelectron spectra as follows. The photoelectron energy distribution $Y(E)$ around the first ATI peak is fitted by a Gaussian, 
\begin{equation}
    Y(E) = A\exp{\left(-\frac{(E- E_C)^2}{\text{w}^{2}}\right)}+Y_0.
\end{equation}
We note that, especially for high intensities, the shape of the ATI peaks deviates significantly from a Gaussian shape. Nevertheless, we obtain good agreement for the peak positions, see figure ref{figs:consistency}, which shows the measured intensity-dependent spectra along with the retrieved centroids of the ATI peaks measured at each laser power and target. The procedure is repeated for 25 different values of the measured laser power, which yields the intensity-dependence of the ATI peak positions displayed in Fig.~\ref{fig4:peakshift}. 

Figure \ref{fig4:peakshift} shows the offsets $E_0 = E_C-2.4\eV$ of the ATI comb, retrieved by the fitting procedure described above, as a function of the calibrated laser intensity. Indeed, the ATI peaks measured for Ar and Ar-\HHO exhibit significantly different shifts as a function of intensity. In contrast, the intensity-dependence of the ATI peak positions measured for Ar and Ar$_2$ are nearly identical. Indeed, the ionization potentials of Ar ($15.76\eV$) and Ar$_2$ ($15.65\eV$) are very close and their ATI peaks exhibit only a small offset with a mean of $(0.08 \pm 0.01)\eV$, in good agreement with the difference in ionization potential. 
Analogously, the energy offset between the ATI peaks measured for \HHO and Ar-\HHO remains essentially constant at $0.26 \eV$ within a margin of $\pm 0.06 \eV$ over the entire intensity range studied in the present experiment. Therefore, it is reasonable to assume that the peak offset corresponds to the difference in ionization potential (cf.~equation \ref{eq:ATI_peak}). With the known ionization potential of \HHO of $12.6\eV$, we find a value of $(12.4 \pm 0.1)\eV$ for Ar-\HHO, where the uncertainty corresponds to the statistical error of the differences in the measured peak positions. We point out that this deviation is independent of the intensity determination conducted above.

We will now further investigate the differences in the intensity-dependence of the ATI peak positions recorded for different species. Clearly, the red-shift of the ATI peaks (cf.~equation \ref{eq:ATI_peak}) is target-dependent. Specifically, the linear fit functions, applied over the entire intensity range of the experiment, yield a slope of $-2.2\eV / (10^{14}\Wcm)$ for Ar and Ar$_2$. These values are in reasonable agreement with the ponderomotive shift of $-2.5\eV / (10^{14} \Wcm)$  for 515-nm light and the difference is consistent with the effect of focal volume averaging. The TDSE results for Ar exhibits some deviations from linearity, likely due to the effect of resonances. Up to an intensity of approximately $1.0 \cdot 10^{14} \Wcm$, the intensity-dependent peak shift of  the TDSE results agrees with the experimental results, however, the curve is blue shifted by approximately 0.2\,eV. Such deviation is possible despite the intensity calibration because the fitting performed above is carried out for the positions of ATI orders 1 to 10, while in Fig.~\ref{fig4:peakshift}, only ATI order 1 is analyzed. Above $1.0 \cdot 10^{14} \Wcm$, the measured and calculated peak positions deviate, potentially because the influence of resonances is exaggerated in the one-dimensional TDSE calculations. This observation emphasizes that the linear relationship between ATI peak position and intensity is only applicable in the absence of resonances.  

The ATI peak shift measured for Ar-\HHO  and \HHO  is only $-1.5\eV / (10^{14}\Wcm)$, significantly less than  the ponderomotive shift of $-2.5\eV / (10^{14} \Wcm)$  for 515-nm light. A first candidate to explain the observed differences between Ar and Ar$_2$ on one hand and \HHO and Ar-\HHO on the other hand is the AC stark shift of the ground state. 
However, the ground state polarizabilities of argon ($\alpha_\textrm{Ar} = \SI{11.07}{a_0^3}$) and water ($\alpha_\textrm{H2O} = \SI{10.74}{a_0^3}$) \cite{Ge2017, Gaiser2018, Lesiuk2023} are very similar and both lead to a Stark shift of the ground state energy by $-0.2\eV / 10^{14}\,\Wcm$. Hence, the difference of ground state polarizabilities for argon and water is too small to explain the observed difference in the intensity-dependent ATI peak shifts of the two targets. Furthermore, as the different shift is observed over the entire intensity range, resonances are unlikely to be responsible for the observed differences. We propose that the smaller red shift of the ATI peaks observed for \HHO and Ar-\HHO results from the permanent dipole \cite{Delone1999}. Its effect may become measurable due to the orientation-dependence of the ionization probability of water \cite{Picca2012}. However, a quantitative calculation of the intensity-dependent ATI peak shift of water is outside the scope of the present paper. 

\subsection{Rescattering from Ar-\HHO}

\begin{figure}
    \centering
    \includegraphics[width = 0.5 \textwidth]{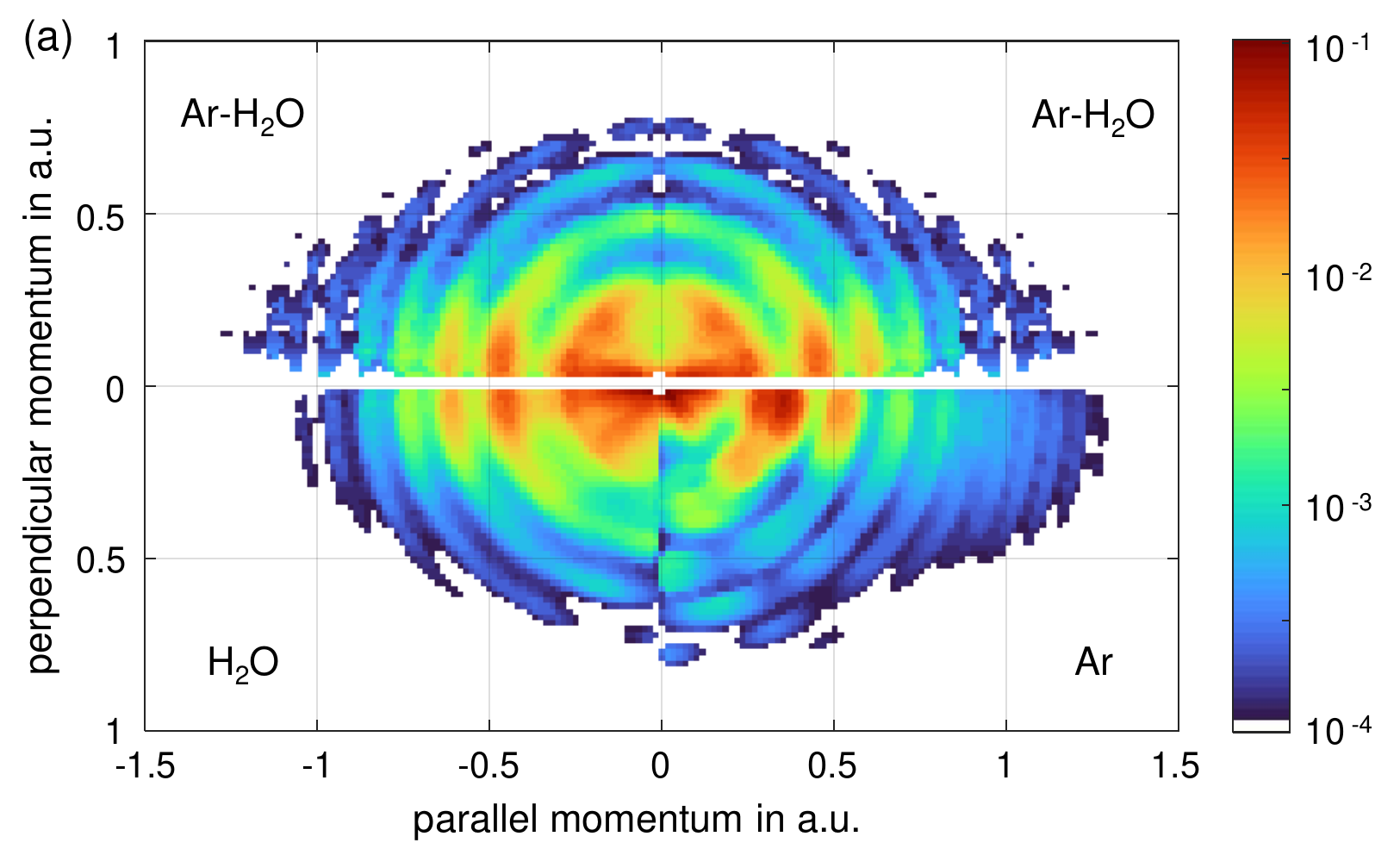}

    \caption{\small \justifying (a)Photoelectron momentum distribution recorded for non-dissociative single-ionization of Ar-\HHO (upper half), compared to the ones recorded for \HHO (bottom left), and Ar (bottom right). The ionization yield is integrated over all intensity values and normalized to 1 by dividing by the number of detected events for each species. The plot shows the normalized ionization yield per momentum bin, as a function of the momentum components parallel $p_{||} = p_z$ and perpendicular $p_\perp = \sqrt{p_x^2+p_y^2}$ to the laser polarization, taking into account the appropriate Jacobian. The minimum of the color scale is limited by the low statistics of the Ar-\HHO channel.}
    \label{fig6:momentum distributions}
\end{figure}

In this section, we return to the observation of enhanced backscattering from Ar-\HHO with respect to \HHO. Figure \ref{fig6:momentum distributions} presents the photoelectron momentum distributions measured for Ar-\HHO (upper half), \HHO (bottom left quadrant) and Ar (bottom right quadrant) corresponding to the ATI spectra presented in Fig.~\ref{fig2:ATI} at high intensity. The momentum distribution measured for Ar$_2$ is compared to the one for Ar in Fig.~\ref{figA:Ar2_Ar}. The central, low-energy, part of the momentum distributions for all three targets are governed by direct electron emission and full ATI rings are observed. At energies significantly larger than $2\,U_\mathrm{P}$, corresponding to $p \approx \SI{0.6}{\au}$, re-scattered electrons dominate and the angular distributions are significantly narrower such that only segments of the ATI rings are observed. 

We first compare Ar-\HHO to \HHO and observe that the low-energy part of the momentum distribution is nearly identical for the two targets. The slightly lower ionization potential of Ar-\HHO leads to slightly larger radii of the ATI rings.  In the high-energy part along the laser polarization, the signal observed for Ar-\HHO is slightly larger than for \HHO. This is evident for the highest ATI peaks but also the signal in the intermediate ones is higher by a a factor of $\approx 2$, see Fig.~\ref{fig2:ATI}. This enhancement suggests a larger contribution from backscattering in the case of Ar-\HHO. Compared to Ar (bottom right quadrant), however, the signal remains slightly lower. The stronger signal observed for Ar might be a consequence of focal volume averaging, which leads to higher effective intensity for Ar, and therefore higher signal at larger electron momenta. Our results suggest that in Ar-\HHO, the relatively large Ar atom acts as a scattering center, which enhances the contribution of backscattering to the photoelectron yield. Indeed, for the differential elastic electron scattering cross section around $180^\circ$, two to three times larger values have been reported for Ar \cite{Bell1984} than for water \cite{Machado1995}, consistent with our results. 

\section{Summary and Outlook}

In conclusion, we have presented measurements of ATI of Ar-\HHO and compared them to simultaneous measurements of both Ar and \HHO. We found that the peak shift for Ar-\HHO matches that for \HHO, while the ionization potential is lowered by some $0.2\,\eV$ with respect to \HHO. These observations indicate that, on the one hand, strong-field ionization occurs on the \HHO side of the weakly bound Van-der-Waals molecule. On the other hand, the reduction of the ionization potential indicates that the molecular ion is more tightly bound than neutral Ar-\HHO. This raises the question about details of the potential energy landscape of the Ar-\HHO cation, for example where the argon atom resides with respect to the H atoms. A suitable tool to shed light on this question is laser-driven Coulomb explosion imaging. 
While the Ar atom has little effect on the initial step of strong-field ionization, we have further shown that elastic rescattering in Ar-\HHO is enhanced by the presence of the Ar atom. By extension to inelastic rescattering, this suggests the existence of a special mechanism for non-sequential double ionization, where a first electron is emitted from water and its subsequent recollision with the parent ion leads to ionization of the argon atom. This mechanism could lead to a peculiar effect: at relatively low intensity and sufficiently long wavelength, Ar-\HHO could be doubly ionized even before single ionization of Ar becomes efficient.

\section*{Acknowledgements}
We thank T. Weber, A. Rose, H. W\"ohl, F. Ronneberger, T. Farhadova, J. Yu, S. Voss and A. Czasch for technical support, in particular when setting up the COLTRIMS. Fruitful discussions with M. Lesiuk and R. Della Picca are acknowledged. This work has been funded by the DFG under Project No. 437321733. 

\bibliography{ar-water}

\newpage
\appendix

\section{Intensity-dependent ATI peak positions}
\begin{figure} [ht]
    \begin{subfigure}[b]{0.235\textwidth}
  	\includegraphics[width = \textwidth]{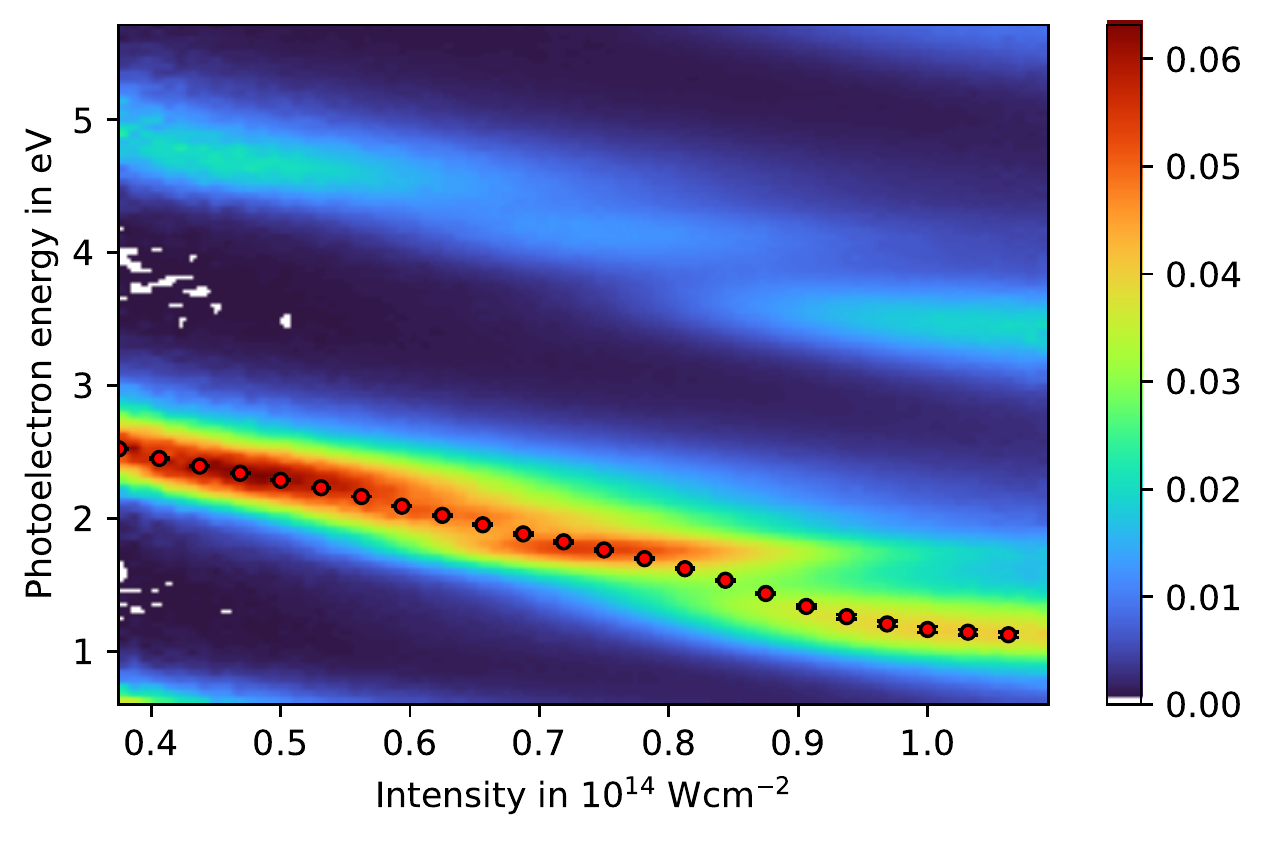}
        \caption{}
    \end{subfigure}
    \begin{subfigure}[b]{0.235\textwidth}
  	\includegraphics[width = \textwidth]{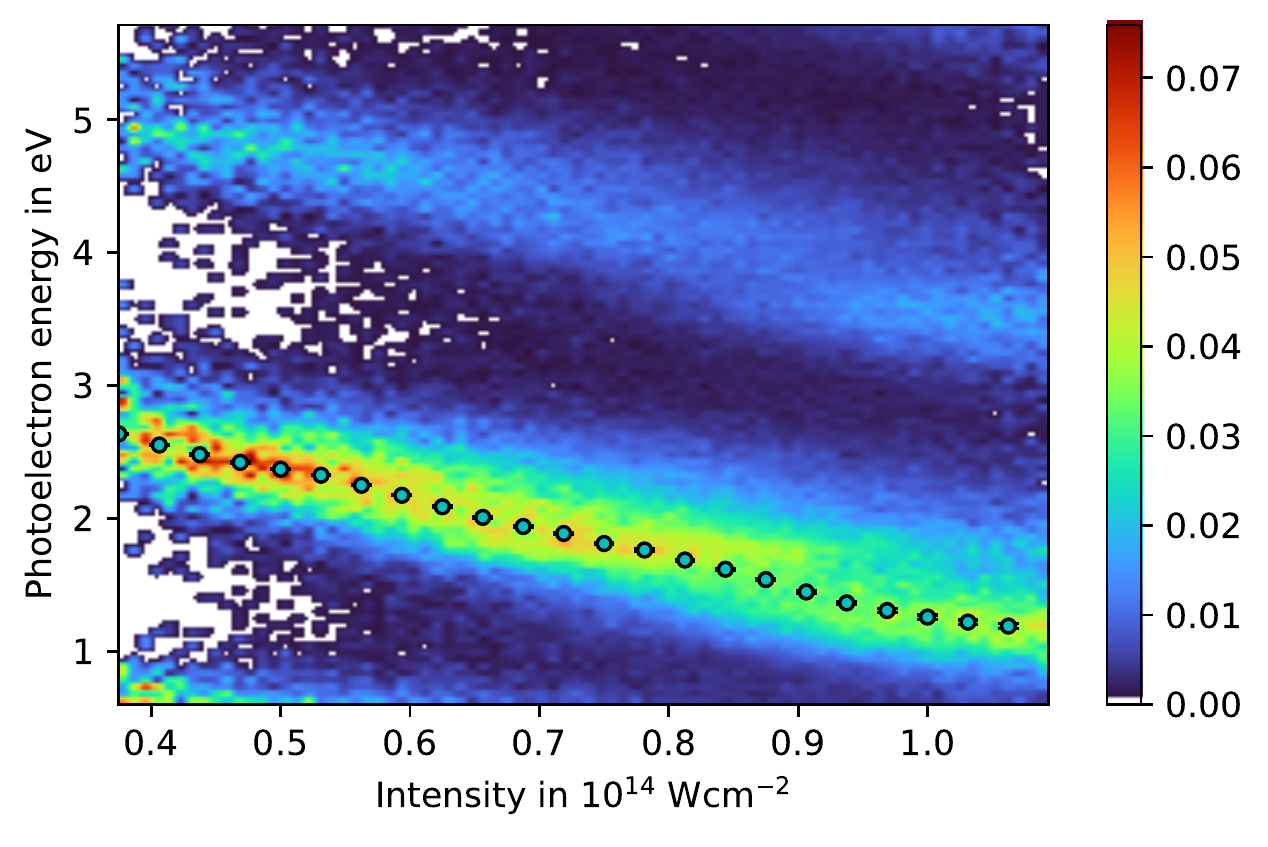}
        \caption{}
    \end{subfigure}
    \begin{subfigure}[b]{0.235\textwidth}
  	\includegraphics[width = \textwidth]{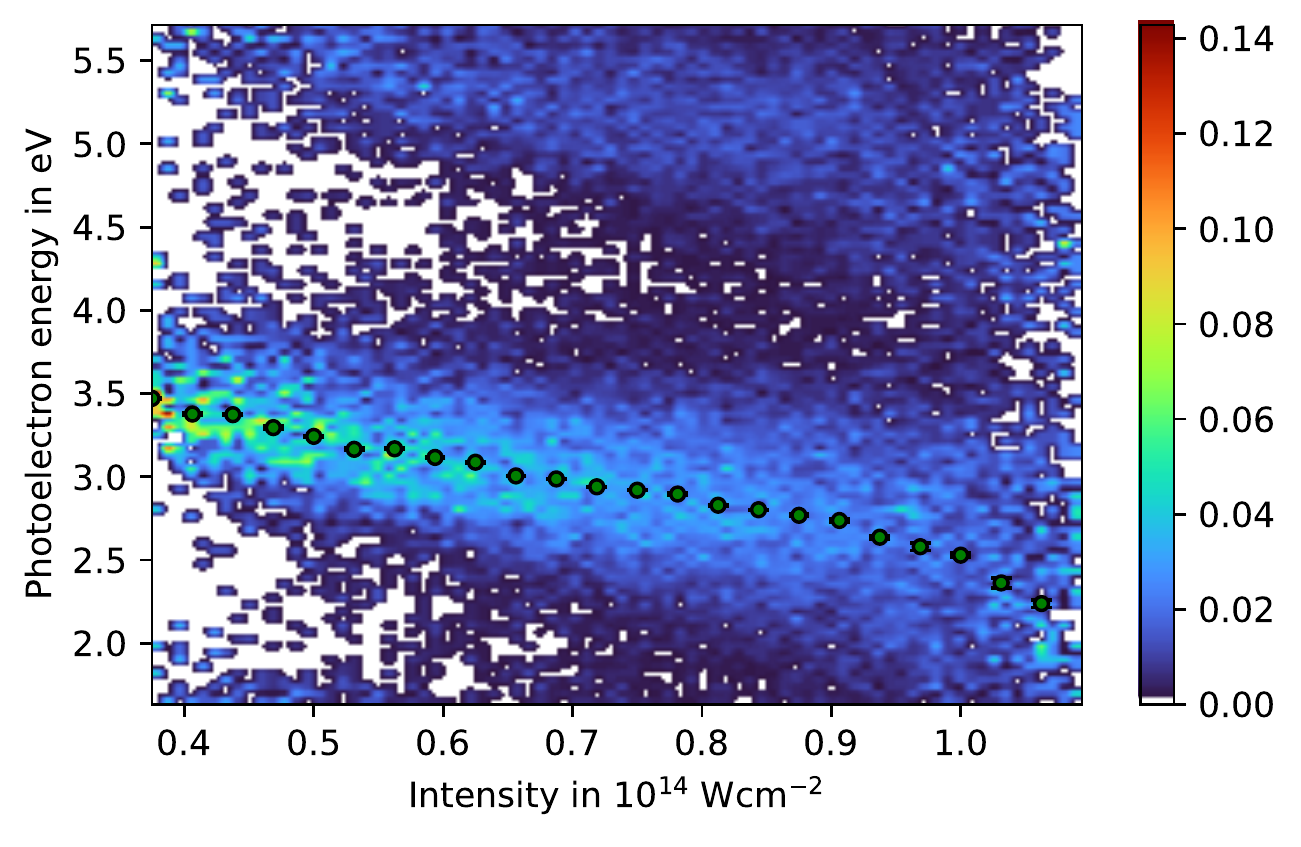}
        \caption{}
    \end{subfigure}
    \begin{subfigure}[b]{0.235\textwidth}
  	\includegraphics[width = \textwidth]{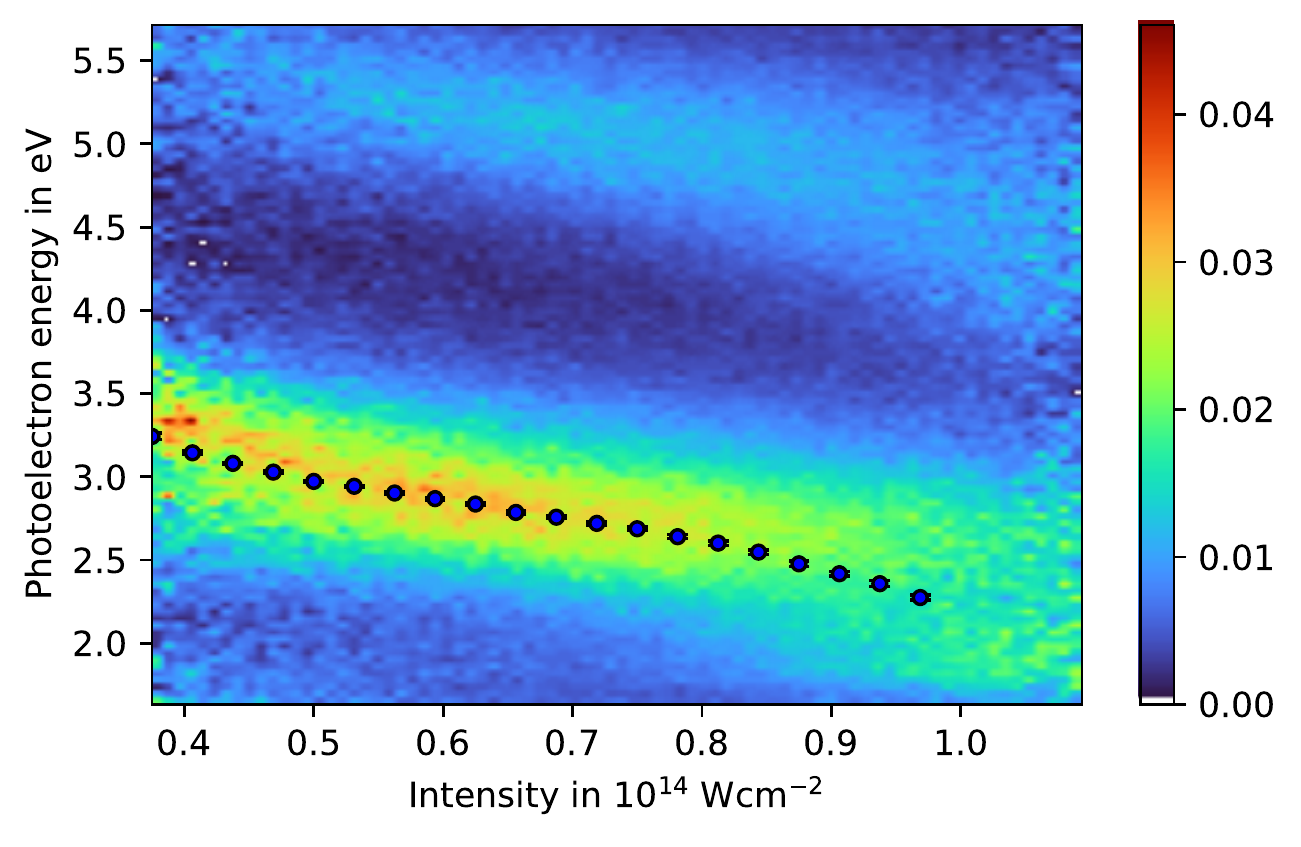}
        \caption{}
    \end{subfigure}
    \caption{\small \justifying Intensity-dependent ATI spectra of (a) argon, (b) argon dimer, (c) argon-water and (d) water. The dots mark the centroids of Gaussians fitted to the experimental data at various intensity values. The statistcal error of the fits is on the order of the dot size.}
    \label{figs:consistency}
\end{figure}

Fig.~\ref{figs:consistency} shows experimental results for the four targets along with the centroids obtained from the fitting procedure described above. The fitting is performed in a range of $\pm\SI{1}{eV}$ around the approximate position of the first ATI peak. The accuracy of the results is not notably improved if the fitting is repeated to ATI peaks of higher order. The zeroth ATI peak was ignored because of channel closing. In the case of water, no accurate fit was obtained at the highest intensities used in the experiment. 

\section{Photoelectron momentum distribution for argon dimers}
\begin{figure} [ht]
    \centering
    \includegraphics[width = 0.45 \textwidth]{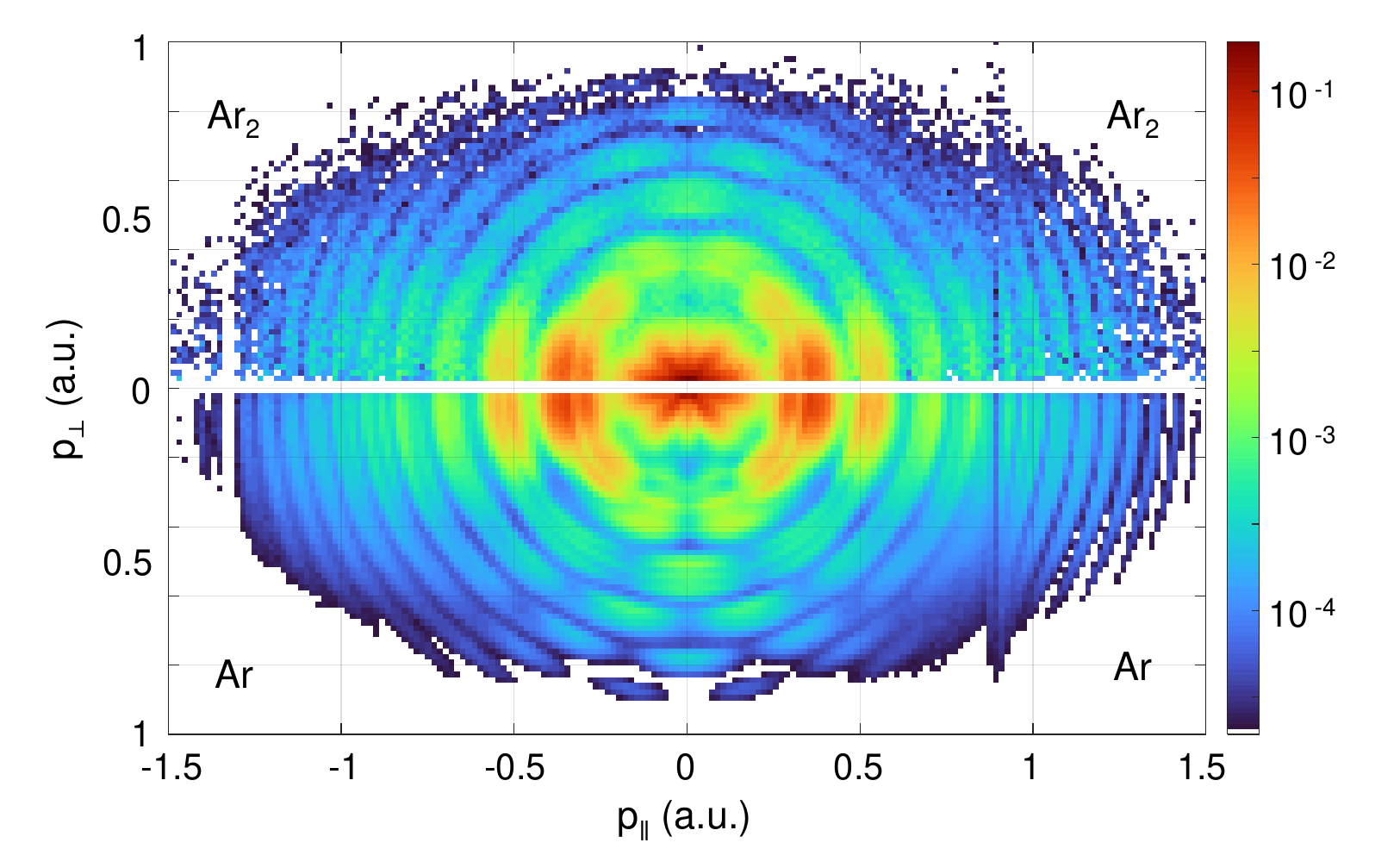}
    \caption{\small \justifying Same as Fig.~\ref{fig6:momentum distributions} but shown the momentum distribution recorded for non-dissociative single-ionization of Ar$_2$ (upper half) compared to the one recorded for Ar (bottom half), as in Fig.~\ref{fig6:momentum distributions}. At $p_{||}\approx -1.3\au$ and $p_{||}\approx 0.8\au$, the detector provides no resolution for the perpendicular momentum.}
    \label{figA:Ar2_Ar}
\end{figure}

Figure \ref{figA:Ar2_Ar} compares the photoelectron momentum distributions recorded for non-dissociative single ionization of Ar$_2$ to that for Ar. The momentum distributions for the two targets are very similar to each other, in agreement with earlier experiments using 800\,nm light \cite{VonVeltheim2013}. A slight yield enhancement at minimal electron energy reported for the dimer in Ref.~\cite{VonVeltheim2013} can be identified.

\end{document}